\documentclass[
superscriptaddress,
preprint,
amsmath,amssymb,
aps,
floatfix,
]{revtex4-1}

\usepackage{graphicx}% Include figure files
\usepackage{dcolumn}% Align table columns on decimal point
\usepackage{bm}% bold math
\usepackage[mathlines]{lineno}% Enable numbering of text and display math
\begin{document}

%\begin{CJK*}{GBK}{song}
%\preprint{APS/123-QED}
\title{Propagation of an Airy-Gaussian beam in defected photonic lattices}
\author{Zhiwei Shi}
\thanks{Corresponding author: szwstar@gdut.edu.cn}
\affiliation{School of Electro-mechanical Engineering, Guangdong University of Technology, Guangzhou 510006,P.R.China}
\author{Jing Xue}
\affiliation{School of Information Engineering, Guangdong University of Technology, Guangzhou 510006,P.R.China}
\author{Xing Zhu}
\affiliation{Department of Physics, Guangdong University of Education , Guangzhou 510303, China}
\author{Yang Li}
\affiliation{School of Electro-mechanical Engineering, Guangdong University of Technology, Guangzhou 510006,P.R.China}
\author{Huagang Li}
\thanks{Corresponding author: lihuagang@gdei.edu.cn}
\affiliation{Department of Physics, Guangdong University of Education , Guangzhou 510303, China}
%\address{$^1$Faculty of Information Engineering, Guangdong University of Technology, Guangzhou 510006,P.R.China}
%generates the title
%\maketitle
%insert the table of contents
%\tableofcontents
%\section{Start}
\begin{abstract}
%% Text of abstract
We investigate numerically that a finite Airy-Gaussian (AiG) beam varies its trajectory and shape in the defected photonic lattices. The propagation properties and
beam self-bending are controlled with modulation depth and period of the photonic lattices, positive and negative defects, beam distribution factor and nonlinearity change. For positive defects, the pseudo-period oscillation and localization of the AiG beam may be formed under a certain condition, while the beam is diffused for negative defects. Moreover, the solitons may appear during the propagation process when the self-focusing nonlinearity is introduced.
\end{abstract}
%\ocis{(190.0190) Nonlinear optics; (190.6135) Spatial solitons; (350.5500) Propagation}
%\begin{keyword}
%Parity-time \sep Saturable nonlinearity  \sep Discrete  multipole dark solitons
%\end{keyword}

\maketitle
%\end{frontmatter}
\section{Introduction}

In 1979, the Airy quantum wave packet of infinite extent was introduced by Berry and Balazs as free particle solutions of the Schr\"{o}dinger equation~\cite{1}. This new kind of nondiffracting beams can be generated
by a Gaussian beam through the cubic phase modulation and Fourier transform lens~\cite{2,3,4}. These beams are self-bending and nondiffracting during propagation. More interestingly, they have the intrivial property of self-healing after being obscured by an obstacle placed in their propagation pathes~\cite{5}. In practice, one knows that the initial Airy beam containing infinite energy is not realizable. For solving this problem, Siviloglou and Christodoulides introduced a finite-energy initial Airy beam to extend the model and found that these beams still exhibit the property of self-bending in the propagation process~\cite{2}. Further, Siviloglou et al. confirmed experimentally the existence of the finite-energy Airy beams at first~\cite{3}. Based on these cases, the Airy beam had applied widely, such as optical clearing of microparticles~\cite{6},
versatile linear bullets~\cite{7}, curved plasma channel generation~\cite{8,9}, and vacuum electron acceleration~\cite{10}.

We know that photonic lattice structures can dramatically change the propagation dynamics of light. So, when one lattice structure is altered, the propagation properties of optical Airy beams should be affected.
Recently, Lu\v{c}i\'{c} \textsl{et al} reported that a finite Airy beam changes its trajectory and shape in optically induced waveguide arrays consisting of different kinds of defects~\cite{11}. Moreover, the propagation of Airy beams inside a two-dimensional optically induced photonic lattice had been numerically studied with an isotropic refractive index potential~\cite{12} and investigated both theoretically and
experimentally with the lattices fabricated by optical induction in photorefractive strontium barium niobate~\cite{13}. Airy beam potential application in nonlinear optics regimes had also been researched in the past few years.
The dynamics of Airy beams propagating from a nonlinear medium to a linear medium was studied~\cite{14}. Formation of self-trapped accelerating optical
beams was demonstrated with different self-focusing nonlinearities~\cite{15}, ranging from Kerr and saturable to quadratic~\cite{16,17}, and also with an optically induced refractive-index gradient~\cite{18}.

As a generalized form of the Airy beams, AiG beams describe in a more realistic way the propagation of the Airy beams because AiG beams carry finite energy and maintain the diffraction-free propagation properties within a finite propagation distance~\cite{19}. The AiG beams had been studied both theoretically and experimentally by many researchers~\cite{19,20,21,22,23}. Deng et al. had theoretically studied the propagation of the AiG beam in nonlinear media~\cite{20,21}. They found that the self-acceleration effect of the AiG beam becomes weaker as the distribution factor increases. When the initial input power increases, the quasi-breather is finally observed. Ez-Zariy~\cite{22} and Zhou~\cite{23} et al. had discussed AiG beams passing through a misaligned optical system with finite aperture and in the fractional Fourier transform plane, respectively. Similarly asymptotic preservation of an accelerating property was observed with AiG beams in nonlinear media and with Airy beam introduced in defected waveguide arrays. This gives us motivation to study the impact of defects in photonic lattices on AiG beams.

In this paper, we numerically study the propagation dynamics of AiG beams in two-dimensional photonic lattices including defects. We realize different defect lattices by embedding the positive and negative defects into the regular lattice and research the influence onto the AiG beam. The defects amazingly change the beam propagation dynamics. For
the negative defect the beams undergo a strong diffusion,
while in the presence of the positive defect they can be compressed and may form localized waves. Additionally, the propagation dynamics and beam
acceleration are controlled by varying nonlinearity. For the self-focusing nonlinearity, the beam may form optical solitons during the propagation process.

\section{The theoretical background and numerical results}

To study the propagation characteristics of AiG beams in defected photonic lattices, along the propagation distance $z$, we consider the nonlinear Schr\"{o}dinger equation
\begin{equation}
i\frac{\partial E}{\partial Z}+\frac{1}{2}(\frac{\partial^2E}{\partial X^2}+\frac{\partial^2E}{\partial Y^2})+V(X,Y)E+\gamma|E|^2E=0,
\label{eq:one}
\end{equation}

\begin{figure}[htbp]
\centering
\fbox{\includegraphics[width=\linewidth]{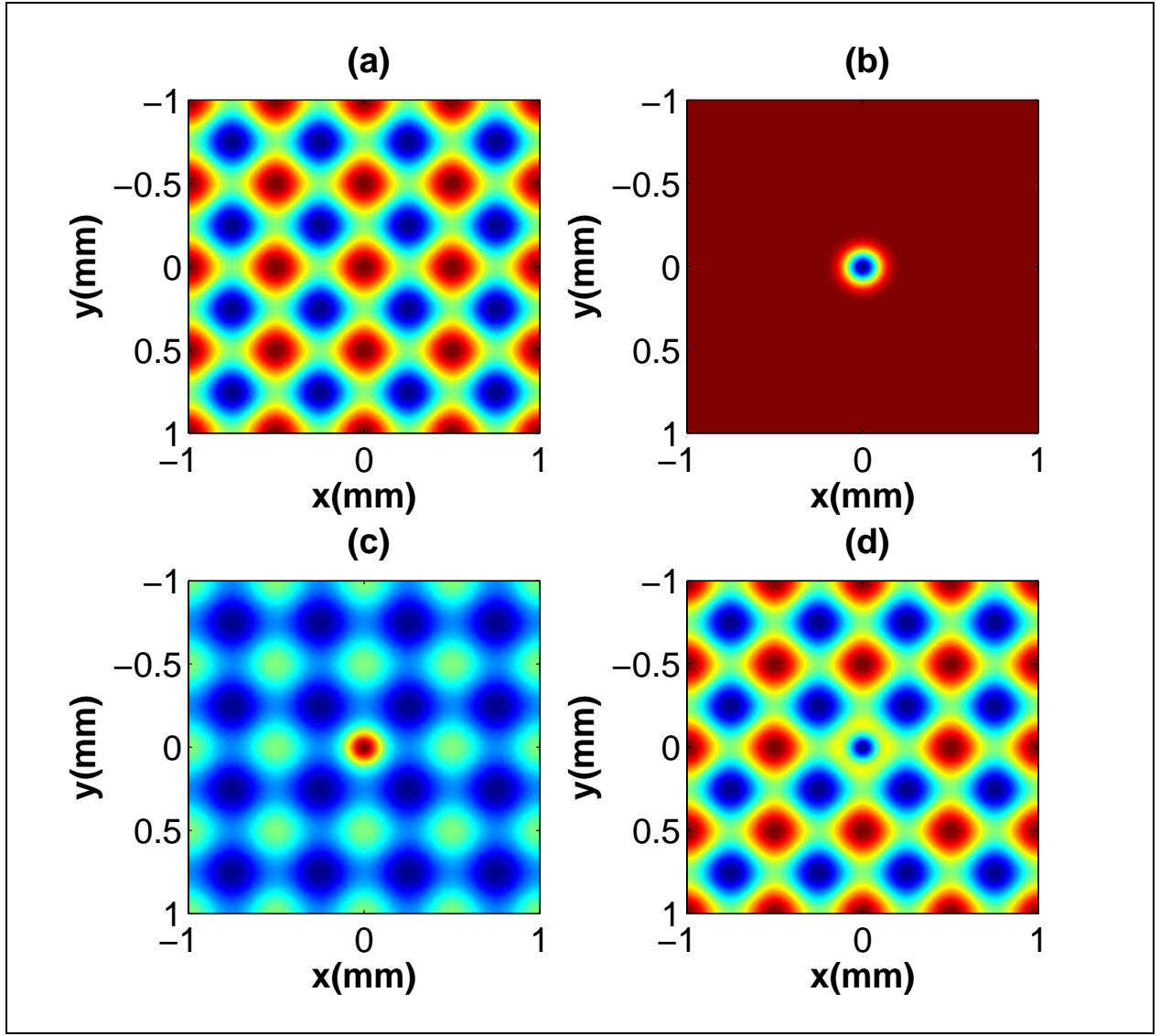}}
\caption{(Color online) Defect generation in optical photonic lattice. (a) The regular lattice distribution $(\delta n=0)$. (b) The Gaussian beam intensity distribution. Numerical realization of (c) positive ($\delta n=1$)and (d) negative ($\delta n=-1$) defect lattices.}
\label{fig:one}
\end{figure}
where $E$ is a slowly varying envelope, $X=x/w_0$ and $Y=y/w_0$ are the dimensionless transverse coordinates scaled by the characteristic length $w_0$,
$Z=z/kw^2_0$ with $k=2\pi/\lambda$, $V(X,Y)=A_n(cos^2(\pi Xw_0/d)+cos^2(\pi Yw_0/d)(1+\delta n\exp(-(X^2+Y^2)))$ is the periodic refractive-index profile of the array with the lattice modulation period $d$ and depth $A_n$ and defect depth $\delta n$. The nonlinearity coefficient $\gamma=\pm1$ denotes self-focusing and self-defocusing nonlinearity, respectively. Here, we assume $w_0=100\mu m$ and the wave length $\lambda=500nm$. Figure \ref{fig:one} shows the basic scheme of the defect realization
using a Gaussian beam $\delta n\exp(-(X^2+Y^2))$. The regular lattice distribution and Gaussian beam intensity distribution are illustrated in Figs.~\ref{fig:one}(a) and ~\ref{fig:one}(b), respectively.
Figs.~\ref{fig:one}(c) and~\ref{fig:one}(d) show the numerically calculated refractive index modulation results for both the positive and negative defect lattices. Taking $E(X,Y,0)=A_0Ai(X)Ai(Y)\exp(\alpha_1X+\alpha_2Y)\exp(-(X^2+Y^2)\sigma^2)$ as an initial input AiG beam, we will numerically discuss different cases of propagation of AiG beams in defected photonic lattices by use of the
split-step Fourier method. $A_0$ denotes the constant amplitude, $Ai(\cdot)$ is the Airy function, $\alpha_1=\alpha_2=0.01$ in the exponential function is a parameter associated with the truncation of the AiG beams, and $\sigma$ is the parameter controlling
the beam that will tend to the Gaussian beam with a larger value and the Airy beam with a smaller value.

\begin{figure}[htbp]
\centering
\fbox{\includegraphics[width=\linewidth]{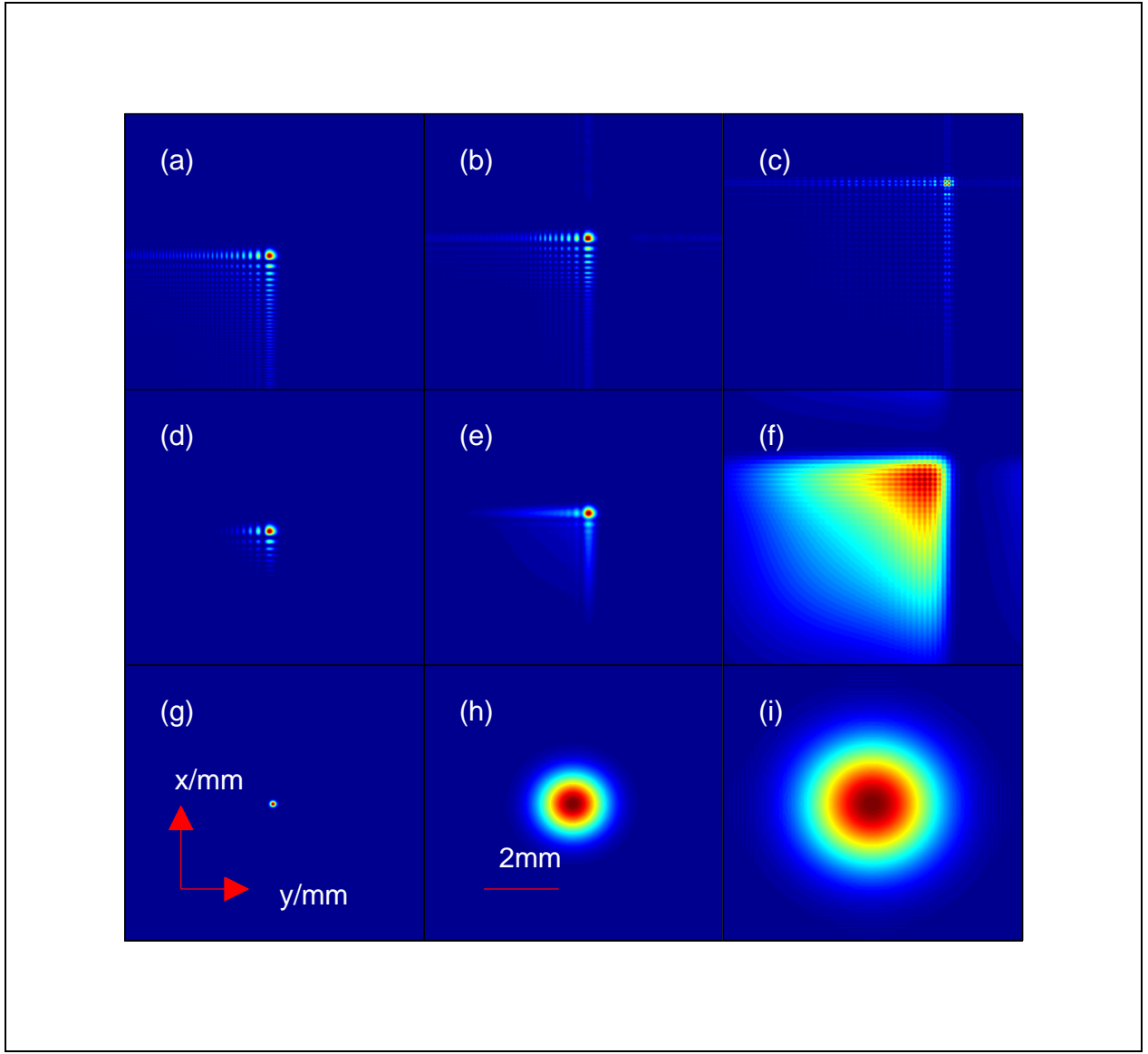}}
\caption{(color online) Intensity profile of AiG beams (a)-(c) $\sigma=0.01$, (d)-(f)
$\sigma=0.1$, (g)-(i) $\sigma=1$, $z$ is chosen as $z=0$m in panels (a), (d), and (g);
$z=0.50$m in panels (b), (e), and (h); $z=1.00$m in panels (c), (f), and (i).}
\label{fig:two}
\end{figure}
Firstly, we discuss that the media is the linear uniform case, that is to say, $\gamma=0$ and $V=0$. Now, we only consider the influence of the adjustable
parameter $\sigma$ on the ratio of Airy beam and Gaussian beam. As mentioned above, when $\sigma$ is smaller, the AiG beam goes to the field distribution of Airy
beams. And, when $\sigma$ is bigger, the AiG beam goes to the field distribution of Gaussian beams. As shown in Figs.~\ref{fig:two} (a)-\ref{fig:two} (c), the field distribution of Airy beams has been exhibited obviously as $\sigma=0.01$. The AiG beam is self-accelerating and diffracting with the increase of transmission distance $z$. When $\sigma=0.1$, there is still properties of an Airy beam for the AiG beam with the shorter propagation distance (see Figs.~\ref{fig:two}(d)-\ref{fig:two} (f)). Moreover, the field distribution of Gaussian beams has been appeared as $\sigma=1$. From Figs.~\ref{fig:two}(g)-\ref{fig:two} (i)), we can see that the AiG beam is self-accelerating no longer, which is very similar to the propagation of the Gaussian beam in the linear uniform media. Thus, we will next discuss the cases for $\sigma=0.1$.

\begin{figure}[htbp]
\centering
\fbox{\includegraphics[width=\linewidth]{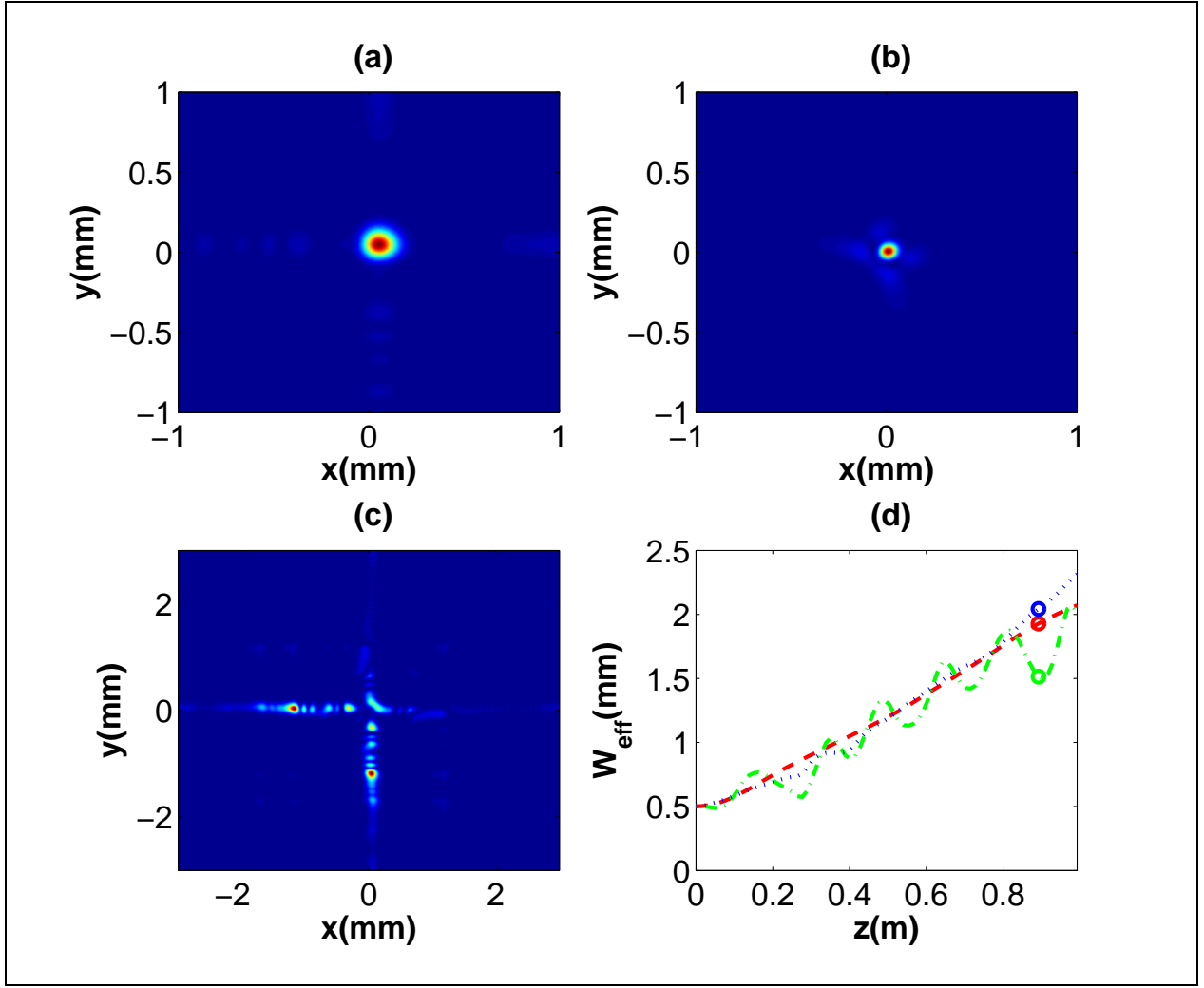}}
\caption{(color online) Intensity profile of AiG beams at $\delta n=0$ (a), $\delta n=1$ (b), and $\delta n=-1$ (c) for $z=900$mm. (d) The change of $W_{eff}$ with $z$. The red dashed line denotes $\delta n=0$; The green dash-dotted line represents $\delta n=1$; The blue dotted line is $\delta n=-1$. Three cycles correspond to (a), (b) and (c). The other physical parameters are $A_0=5$, $A_n=3$, $d=500\mu$m, and $\gamma=0$.}
\label{fig:three}
\end{figure}
\begin{figure}[htbp]
\centering
\fbox{\includegraphics[width=\linewidth]{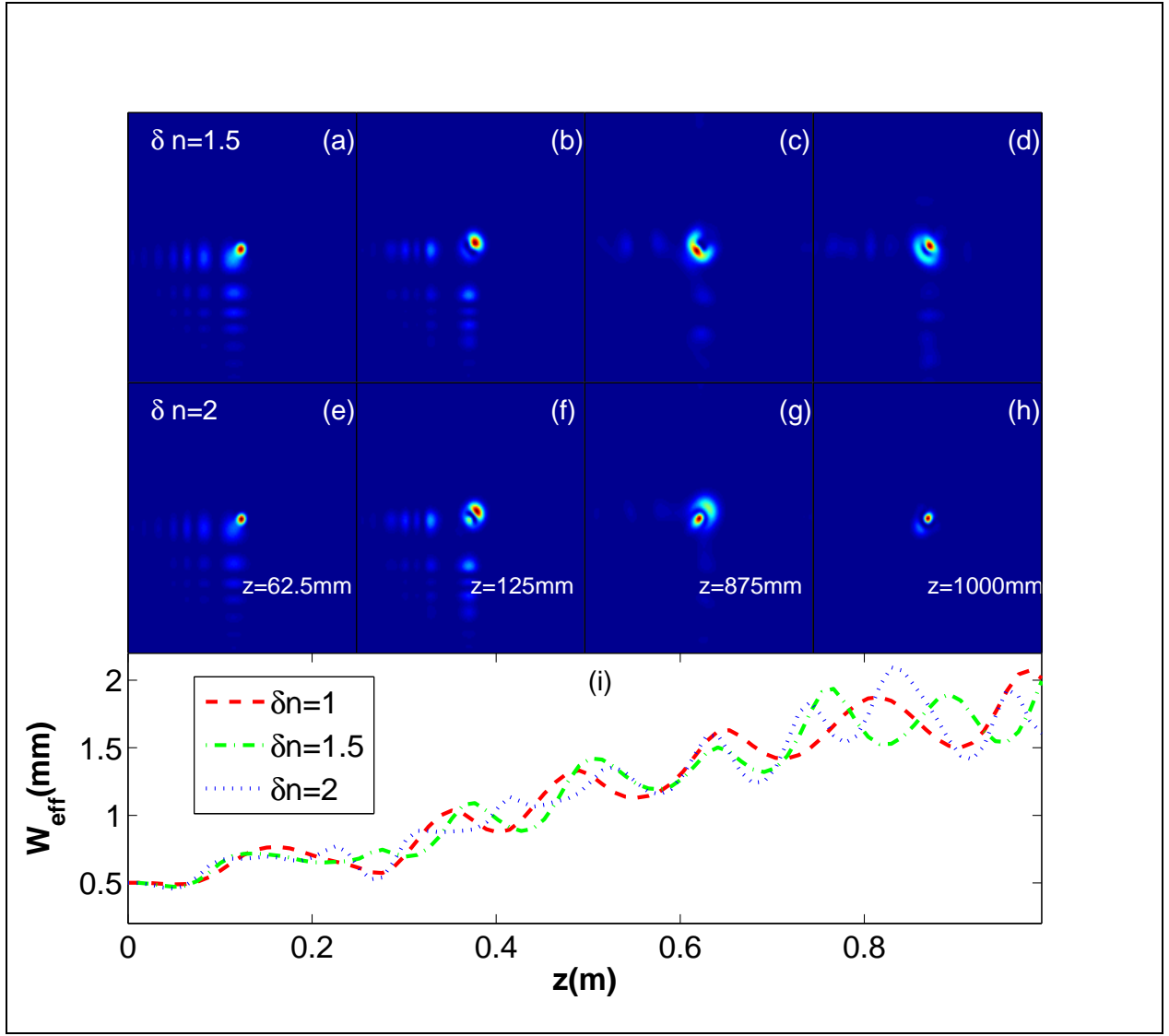}}
\caption{(color online) Intensity profile of AiG beams at $\delta n=1.5$ (a)-(d) and $\delta n=2$ (e)-(h). $z$ is chosen as $z=62.5$mm in panels (a) and (e);
$z=125$mm in panels (b) and (f); $z=875$mm in panels (c) and (g); $z=1000$mm in panels (d) and (h). (i) The change of $W_{eff}$ with $z$. The other physical parameters are same as Fig.~\ref{fig:three}.}
\label{fig:four}
\end{figure}
Secondly, the propagation of AiG beams in the defected photonic lattices are illustrated at $\gamma=0$ as shown in Figure~\ref{fig:three}. Compared with $\delta n=0$ (Fig.~\ref{fig:three} (a)), the AiG beam is compressed in the positive defected lattices at $\delta n=1$ (Fig.~\ref{fig:three} (b)), but one is diffused in the negative defected lattices at $\delta n=-1$ (Fig.~\ref{fig:three}(c)). To further test the influence of the defected lattices, we use approximatively the mean radius of an optical beam $W_{eff}(z)=2(\langle x^2\rangle-\langle x\rangle^2)^{1/2}$~\cite{20} to denote the AiG beam width in the $x$ direction, where $\langle x\rangle=\int ^{+\infty}_{-\infty}x|E(x,0,z)|^2dx/P$ and the beam power $P=\int^{+\infty}_{-\infty}|E(x,0,z)|^2dx$. Fig.~\ref{fig:three}(d) shows $W_{eff}$ versus the propagation distance $z$. We can say that the beams are diffused for $\delta n=0$ and $\delta n=-1$ all the time. However, because of the existence of the negative defect, the beam distribution is different between two cases (see Fig.~\ref{fig:three} (a) and Fig.~\ref{fig:three} (c)). At $\delta n=1$, the beam oscillation emerges though the beam also diffuses. This reason is the ``seesaw battle" between the positive defect in the photonic lattice and the self-accelerating of the beam. The positive defect compresses the beam. On the contrary, the self-accelerating diffuses the beam. The result can be seen from the green dash-dotted line in Fig.~\ref{fig:three} (d). If we further increase the value of $\delta n$, such as $\delta n=1.5$ and $\delta n=2$, the sidelobe of AiG beam will be more compressed as shown in Figs.~\ref{fig:four} (a)-\ref{fig:four} (h). Interestingly, we can see that the beam oscillation pseudo-period will be shorter with the increase of $\delta n$ from Fig.~\ref{fig:four} (i). That is to say, when $\delta n$ is bigger, the beam is compressed faster.

\begin{figure}[htbp]
\centering
\fbox{\includegraphics[width=\linewidth]{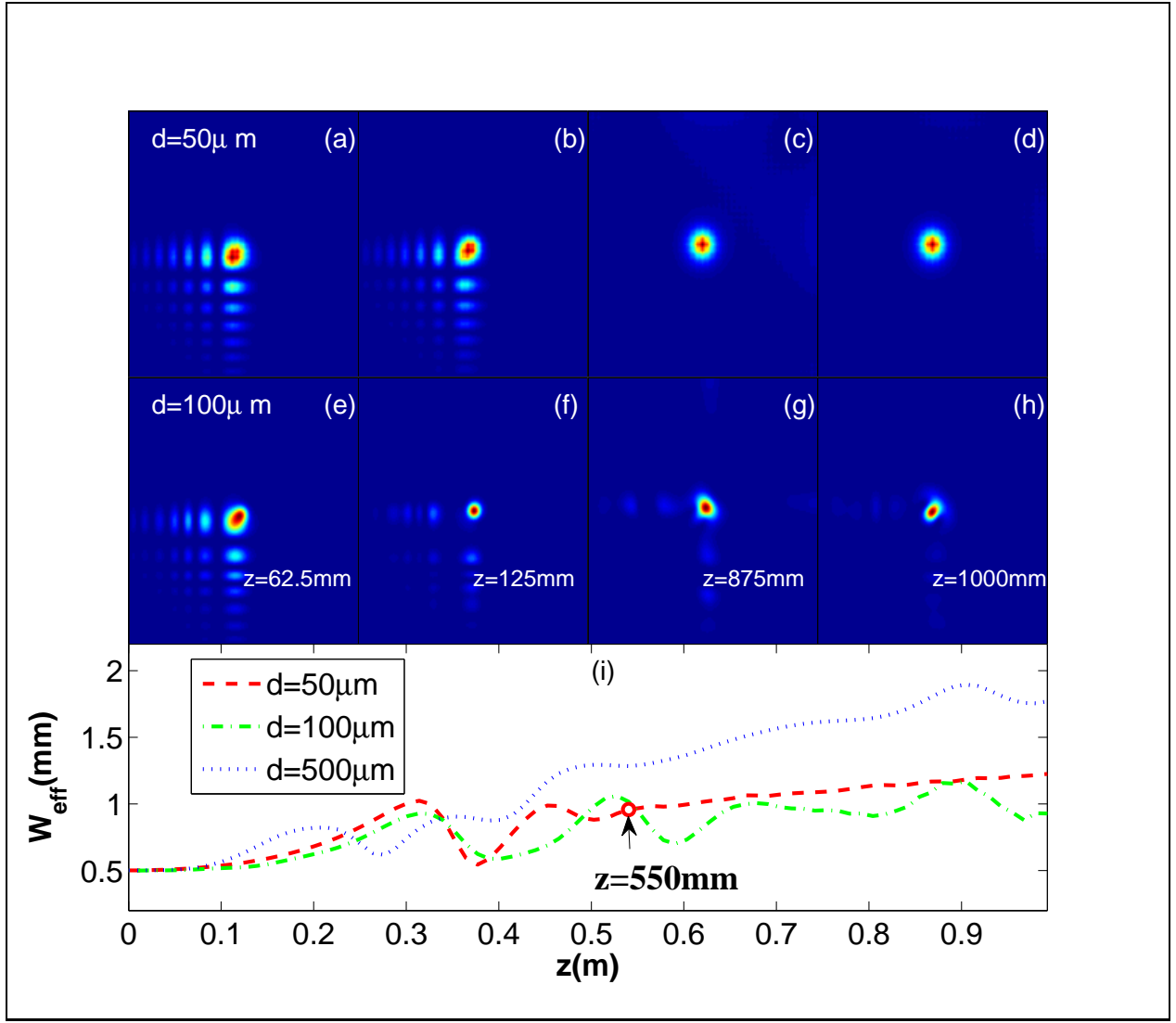}}
\caption{(color online) Intensity profile of AiG beams at $d=50\mu$m (a)-(d) and $d=100\mu$m (e)-(h). $z$ is chosen as $z=62.5$mm in panels (a) and (e);
$z=125$mm in panels (b) and (f); $z=875$mm in panels (c) and (g); $z=1000$mm in panels (d) and (h). (i) The change of $W_{eff}$ with $z$. The other physical parameters are $A_0=5$, $A_n=3$, $\delta n=0.5$, and $\gamma=0$.}
\label{fig:five}
\end{figure}
\begin{figure}[htbp]
\centering
\fbox{\includegraphics[width=\linewidth]{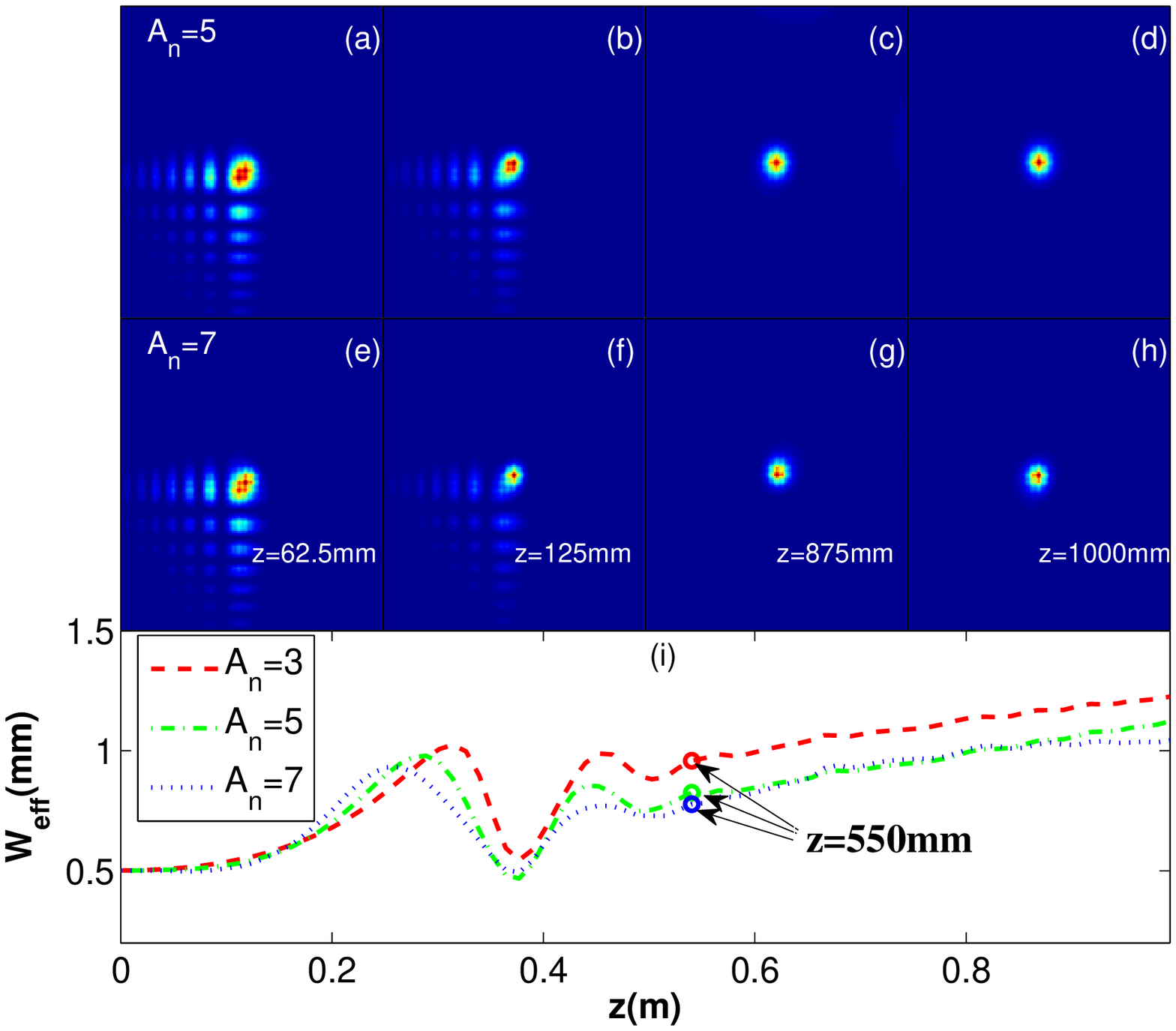}}
\caption{(color online) Intensity profile of AiG beams at $A_n=5$ (a)-(d) and $A_n=7$ (e)-(h). $z$ is chosen as $z=62.5$mm in panels (a) and (e);
$z=125$mm in panels (b) and (f); $z=875$mm in panels (c) and (g); $z=1000$mm in panels (d) and (h). (i) The change of $W_{eff}$ with $z$. The red, green, and blue cycles denote the beam width at $z=550\mu$m. The other physical parameters are $A_0=5$, $d=50\mu$m, $\delta n=0.5$, and $\gamma=0$.}
\label{fig:six}
\end{figure}
Thirdly, we discuss the influence of the modulation period $d$ and depth $A_n$ of the photonic lattices on the propagation properties of AiG beams. At $d=50\mu$m and $d=100\mu$m, the shape of AiG beams is different as shown in Figs.~\ref{fig:five} (a)-\ref{fig:five} (h). Especially,
it is more obvious that the sidelobe of the AiG beams is disappeared at $z=1000$mm (see Figs.~\ref{fig:five} (d) and (h)). The difference of the modulation period
leads to the different structure of lattices, so it influences the AiG beam shape. This is similar to the limitation of aperture on the light beam in geometrical optics. From Fig.~\ref{fig:five} (i), we can find that the effect of the change of modulation period on the beam compression is irregular. However, at $d=50\mu$m, when $z\geq550\mu$m, the beam width doesn't change with $z$ almost. Further, if we fix $d=50\mu$m, the modulation depth increases. We can see that the beams can be more compressed when we observe the beam change for the modulation depth $A_n=5$ and $A_n=7$ from Fig.~\ref{fig:six} (a)-\ref{fig:six} (h). In Fig.~\ref{fig:six} (i), the comparison of $W_{eff}$ among $A_n=3$, $A_n=5$ and $A_n=7$ turns out that they share the similar changing trend. More interestingly, the beam width is almost the same when $z\geq550\mu$m at $A_n=5$ and $A_n=7$. This illustrates that the defect lattices lead to the localization of the AiG beam when the propagation distance is more than a value. And, the beam width can scarcely change with the increase of the modulation depth as the localization of the AiG beam can be formed.

\begin{figure}[htbp]
\centering
\fbox{\includegraphics[width=\linewidth]{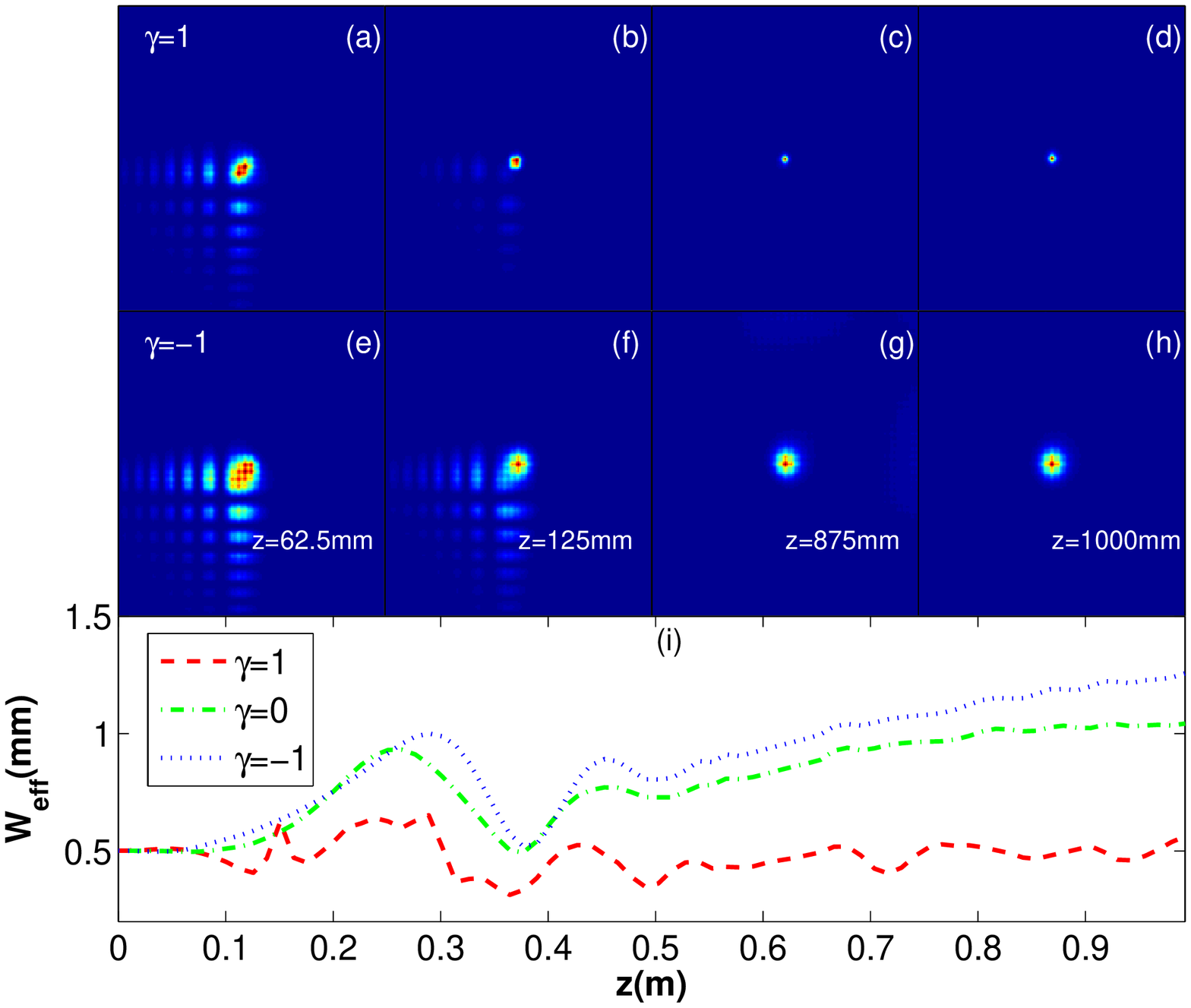}}
\caption{(color online) Intensity profile of AiG beams at $\gamma=1$ (a)-(d), $\gamma=-1$ (e)-(h). $z$ is chosen as $z=62.5$mm in panels (a) and (e);
$z=125$mm in panels (b) and (f); $z=875$mm in panels (c) and (g); $z=1000$mm in panels (d) and (h). (i) The change of $W_{eff}$ with $z$. The red cycle denotes the beam width at $z=550\mu$m. The other physical parameters are $A_0=5$, $d=50\mu$m, $\delta n=0.5$, and $A_n=7$.}
\label{fig:seven}
\end{figure}
\begin{figure}[htbp]
\centering
\fbox{\includegraphics[width=\linewidth]{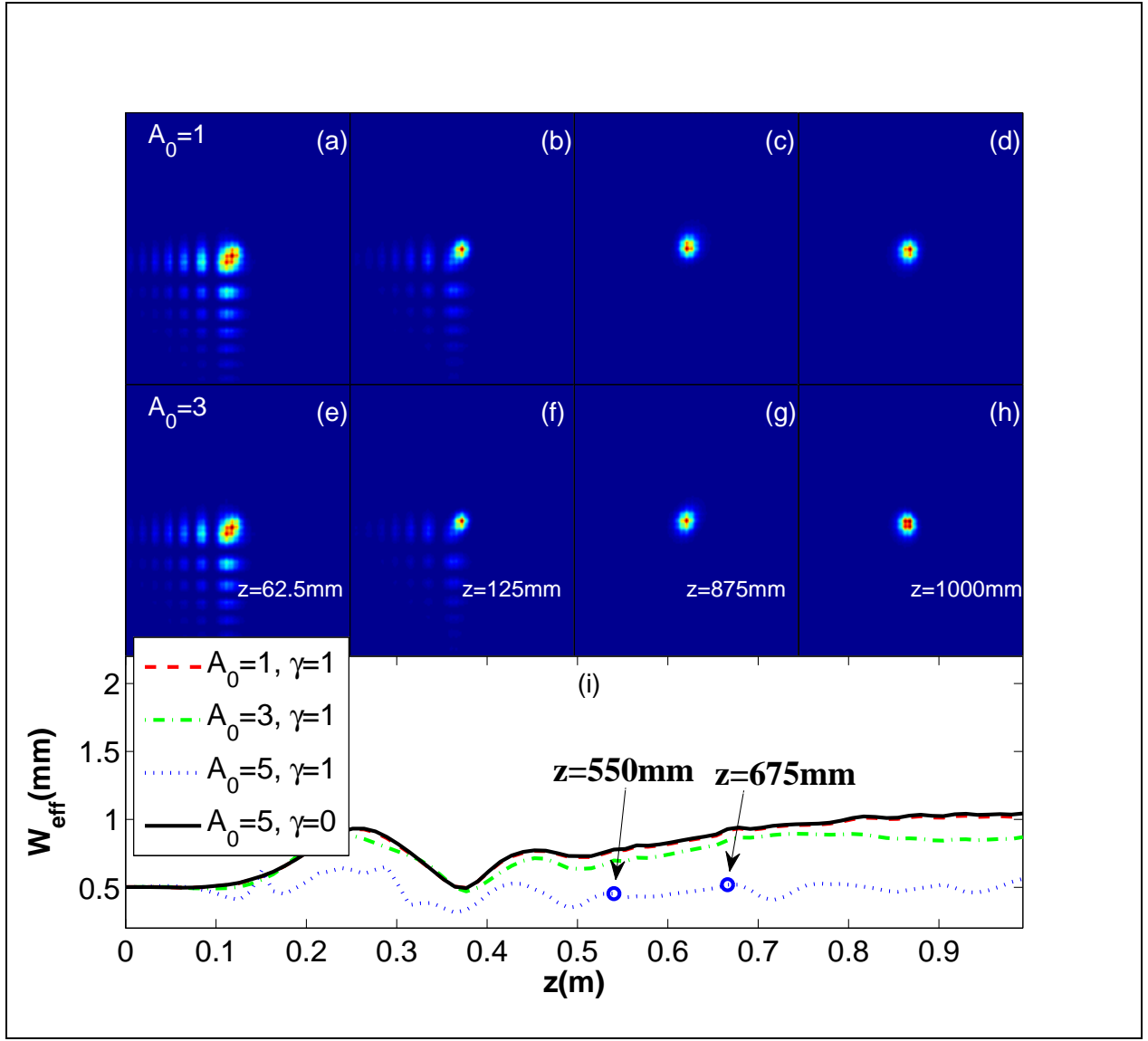}}
\caption{(color online) Intensity profile of AiG beams at $A_0=1$ (a)-(d) and $A_0=3$ (e)-(h). $z$ is chosen as $z=62.5$mm in panels (a) and (e);
$z=125$mm in panels (b) and (f); $z=875$mm in panels (c) and (g); $z=1000$mm in panels (d) and (h). (i) The change of $W_{eff}$ with $z$. The two blue cycles denote the beam width at $z=550\mu$m and $z=550\mu$m, respectively. The other physical parameters are $\gamma=1$, $d=50\mu$m, $\delta n=0.5$, and $A_n=7$.}
\label{fig:eight}
\end{figure}

Finally, the propagation of AiG beams in nonlinear uniform media is considered. From the comparison of beam among $\gamma=0$ (Figs.~\ref{fig:six} (e)-Fig.~\ref{fig:six} (h)), $\gamma=1$ (Figs.~\ref{fig:seven} (a)-Fig.~\ref{fig:seven} (d)) and $\gamma=-1$ (Figs.~\ref{fig:seven} (e)-Fig.~\ref{fig:seven} (h)), we can see that the self-focusing nonlinearity ($\gamma=1$) compresses the beam width, but the self-defocusing one ($\gamma=-1$) diffuses the beam width. Fig.~\ref{fig:seven} (i) further turns out the phenomenon by the change of $W_{eff}$ about the three cases. This shows the nonlinearity works for the beam propagation. To verify the action of the nonlinearity further, we change the input beam power $P$. Here, the power varies through changing the beam amplitude $A_0$. When $A_0$ is bigger, the beam power is bigger. Obviously, one can see that the beam is more severely compressed when the power is bigger from Figs.~\ref{fig:seven} (a)-Fig.~\ref{fig:seven} (d)) and Figs.~\ref{fig:eight} (a)-Fig.~\ref{fig:eight} (h). In Fig.~\ref{fig:eight} (i), we can also get the result. More importantly, at $A_0=5$, the AiG beam may form optical solitons during a certain propagation distance such as from $z=550$mm to $z=675$mm, that is, the beam width doesn't change in the propagation distance. In addition, comparing two cases ($\gamma=0$, $A_0=5$ and $\gamma=1$, $A_0=1$), two lines (black solid line and red dashed line) are almost overlapping. This shows that the action of self-focusing nonlinearity of media on the AiG beam propagation is similar to increasing the beam power.

%The lattice constant W0,can be seen as an effective nonlinearity modulation,
\section{Conclusion}
In conclusion, we have numerically investigated the propagation properties of AiG beams in defected photonic lattices. The depth of
the defect strongly affect the acceleration and the bending of the AiG
beam. The positive defected photonic lattices can compress the beams and lead to the beam pseudo-period oscillation, on the contrary, the negative ones can diffuse the beams. Meanwhile, the modulation period and depth of the lattice can affect the beam propagation properties and may give rise to the beam localization. Moreover, when the nonlinearity is introduced, it also influences the beams. The beam width is compressed in the self-focusing media, but one is diffused in the self-defocusing media. For the different nonlinearity strengthen, the solitons may come into being. Our results can readily be generalized to other kinds of self-accelerated optical beams (e.g. the Airy每Gaussian vortex beam) which are controlled using the presented ideas and methods.

\section*{Acknowledgments}

 This work was supported by the Natural Science Foundation of Guangdong Province of China (Grant No. 2016A030313747) and the National Natural Science Foundation of China (Grant No.11547212).

%\end{CJK*}
%\section{End}

\begin{thebibliography}{99}
\bibitem{1}M.~V.~Berry and N.~L.~Balazs, ``Nonspreading wave-packets," Am. J. Phys. \textbf{47}(3), 264--267 (1979).
\bibitem{2}G. A. Siviloglou and D. N. Christodoulides, ``Accelerating finite energy Airy beams," Opt. Lett. \textbf{32}, 979每-981 (2007).
\bibitem{3}G. A. Siviloglou, J. Broky, A. Dogariu, and D. N. Christodoulides, ``Observation of accelerating Airy beams," Phys. Rev. Lett. \textbf{99}, 213901 (2007).
\bibitem{4}G. A. Siviloglou, J. Broky, A. Dogariu, and D. N. Christodoulides, ``Ballistic dynamics of Airy beams," Opt. Lett. \textbf{33}, 207每-209 (2008).
\bibitem{5}J. Broky, G. A. Siviloglou, A. Dogariu, and D. N. Christodoulides, ``Self-healing properties of optical Airy beams," Opt. Express \textbf{16}, 12880每-12891 (2008).
\bibitem{6}J. Baumgartl, M. Mazilu, and K. Dholakia, ``Optically mediated particle clearing using Airy wavepackets," Nat. Photon. \textbf{2}, 675每-678 (2008).
\bibitem{7}P. Chong, W. Renninger and D. N. Christodoulides, ``Accelerating finite energy Airy beams," Nat. Photon. \textbf{4}, 103--106 (2010).
\bibitem{8}P. Polynkin, M. Kolesik, J. V. Moloney, G. A. Siviloglou, and D. N. Christodoulides, ``Curved Plasma Channel Generation Using Ultraintense Airy Beams," Science \textbf{324}, 229--232 (2009).
\bibitem{9}P. Polynkin, M. Kolesik, and J. V. Moloney, ``Filamentation of femtosecond laser Airy beams in water," Phys Rev Lett. \textbf{103}, 123902 (2009).
\bibitem{10}J. X. Li, W. P. Zhang, and J. G. Tian, ``Vacuum laser-driven acceleration by Airy beams," Opt. Express \textbf{18}, 7300--7306 (2010).
\bibitem{11}N. M. Lu\v{c}i\'{c}, B. M. Boki\'{c}, D. \v{Z}. Gruji\'{c}, D. V. Panteli\'{c}, B. M. Jelenkovi\'{c}, A. Piper, D. M. Jovi\'{c}, and D. V. Timotijevi\'{c}, ``Defect-guided Airy beams in optically induced waveguide arrays," Phys. Rev. A \textbf{88}, 063815 (2013).
\bibitem{12}A. Piper, D. V. Timotijevi\'{c}, and D. M. Jovi\'{c}, ``Acceleration control of Airy beams with optically induced photonic lattices," Phys. Scr. \textbf{T157}, 014023 (2013).
\bibitem{13}F. Diebel, B. M. Boki\'{c}, M. Boguslawski, A. Piper, D. V. Timotijevi\'{c}, D. M. Jovi\'{c}, and C. Denz, ``Control of Airy-beam self-acceleration by photonic lattices," Phys. Rev. A \textbf{90}, 033802 (2014).
\bibitem{14}Y. Hu, S. Huang, P. Zhang, C. B. Lou, J. J. Xu, and Z. G. Chen, ``Persistence and breakdown of Airy beams driven by an initial nonlinearity," Opt. Lett. \textbf{35}, 3952每-3954 (2010).
\bibitem{15}I. Kaminer, M. Segev, and D. N. Christodoulides, ``Self-Accelerating Self-trapped optical beams," Phys. Rev. Lett. \textbf{33}, 207每-209 (2008).
\bibitem{16}A. Lotti, D. Faccio, A. Couairon, D. G. Papazoglou, P. Panagiotopoulos, D. Abdollahpour, and S. Tzortzakis, ``Stationary nonlinear Airy beams," Phys. Rev. A \textbf{84}, 021807(R) (2011).
\bibitem{17}I. Dolev, I. Kaminer, A. Shapira, M. Segev, and A. Arie, ``Experimental observation of self-accelerating beams in quadratic nonlinear media," Phys. Rev. Lett. \textbf{108}, 113903 (2012).
\bibitem{18}Z. Y. Ye, S. Liu, C. B. Lou, P. Zhang, Y. Hu, D. H. Song, J. L. Zhao, and Z. G. Chen, ``Acceleration control of Airy beams with optically induced refractive-index gradient," Opt. Lett. \textbf{36}, 3230每-3232 (2011).
\bibitem{19}M. A. Bandres and J. C. Guti\'{e}rrez-Vega, ``Airy-Gauss beams and their transformation by paraxial optical systems," Opt. Express \textbf{15}, 16719--16728 (2007).
\bibitem{20}D. M. Deng, and H. Li, ``Propagation properties of Airy-Gaussian beams," Appl. Phys. B \textbf{106}, 677--681 (2012).
\bibitem{21}C. D. Chen, B. Chen, X. Peng, and D. M. Deng, ``Propagation of Airy-Gaussian beam in Kerr medium," J. Optics-UK \textbf{17}, 035504 (2015).
\bibitem{22}L. Ez-Zariy, S. Hennani, H. Nebdi, and A. Belafhal, ``Propagation Characteristics of Airy-Gaussian Beams Passing through a Misaligned Optical System with Finite Aperture," Opt. Photon. J. \textbf{4}, 325--336 (2014).
\bibitem{23}Y. M. Zhou, G. Q. Zhou, and G. Y. Ru, ``Properties of Airy-Gauss beams in the fractional fourier transform plane," Prog. Electromagn. Res. \textbf{40}, 143--151 (2014).
%\bibitem{24}L. H. Zhu, Y. H. Nie, and B. D. Lu, ''The concept of the beam width and comparison of its different definitions," Acta Photonica Sinica \textbf{34}, 1476--1479 (2005).
\end{thebibliography}
\end{document}